\documentclass[aps,pra,twocolumn,superscriptaddress,floatfix,
nofootinbib,showpacs,longbibliography]{revtex4-1}
\usepackage[utf8]{inputenc}  
\usepackage[T1]{fontenc}     
\usepackage[british]{babel}  
\usepackage[sc,osf]{mathpazo}
\usepackage[scaled=0.86]{berasans}  
\usepackage[colorlinks=true, citecolor=blue, urlcolor=blue]{hyperref}  
\usepackage{graphicx}
\usepackage[babel]{microtype}  
\usepackage{amsmath,amssymb,amsthm,bm,amsfonts,mathrsfs,bbm} 

\usepackage{xspace}  
\usepackage{pgf,tikz}
\usepackage{xcolor}
\usepackage{multirow}
\usepackage{array}
\usepackage{bigstrut}
\usepackage{braket}
\usepackage{color}
\usepackage{natbib}
\usepackage{multirow}
\usepackage{mathtools}
\usepackage{float}
\usepackage{xcolor,colortbl}
\usepackage{physics}
\usepackage{amsmath}
\usepackage{color}
\usepackage[justification=justified, format=plain]{subcaption}
\usepackage[justification=raggedright]{caption}

\newcommand{\be}{\begin{equation}}
\newcommand{\ee}{\end{equation}}
\newcommand{\ba}{\begin{eqnarray}}
\newcommand{\ea}{\end{eqnarray}}

\def\>{\rangle}
\def\<{\langle}

\begin{document}
	
\title{Assessing non-Markovian dynamics through  moments of the Choi state}	

\author{Bivas Mallick}
\email{bivasqic@gmail.com}
\affiliation{S. N. Bose National Centre for Basic Sciences, Block JD, Sector III, Salt Lake, Kolkata 700 106, India}

\author{Saheli Mukherjee}
\email{mukherjeesaheli95@gmail.com}
\affiliation{S. N. Bose National Centre for Basic Sciences, Block JD, Sector III, Salt Lake, Kolkata 700 106, India}

\author{Ananda G. Maity}
\email{anandamaity289@gmail.com}
\affiliation{Networked Quantum Devices Unit, Okinawa Institute of Science and Technology Graduate University, Onna-son, Okinawa 904-0495, Japan}

\author{A. S. Majumdar}
\email{archan@bose.res.in}
\affiliation{S. N. Bose National Centre for Basic Sciences, Block JD, Sector III, Salt Lake, Kolkata 700 106, India}

\begin{abstract}
 Non-Markovian effects in open quantum system dynamics usually manifest backflow of information from the environment to the system, indicating complete-positive divisibility breaking of the dynamics. We provide a criterion for witnessing such non-Markovian dynamics exhibiting information backflow, based on partial moments of Choi-matrices. The moment condition determined by the positive semi-definiteness of a matrix, does not hold for a Choi-state describing non-Markovian dynamics. We then present some explicit examples in support of our proposed non-Markovianity detection scheme. Finally, a moment based measure of non-Markovianity  for unital dynamics is formulated.

\end{abstract}
\maketitle

\section{Introduction}
According to the postulates of quantum mechanics, closed systems evolve unitarily. However, due to the inevitable interaction with  noisy environments, the system undergoes irreversible phenomena such as dissipation and decoherence.  The theory of open quantum systems provides adequate tools for studying such dynamics comprising of system-environment interactions \citep{alicki,lindblad,gorini,breuer,rivas1,breuerN,alonso, chruscinski2022dynamical}. System-environment interactions are often assumed to be {\it Markovian} where the environment does not keep memory of past interactions with the system and the interaction is considered to be sufficiently weak. However, in realistic scenarios, when the system-environment coupling is not sufficiently weak and the environment has some finite memory,  the description of open quantum systems by the Markovian model may fall short leading to the requirement of the {\it non-Markovian} paradigm \citep{RHP,BLP,blp1,bellomo,arend}. Unlike  Markovian dynamics (i.e, the dynamics without memory effects), non-Markovian dynamics usually contains a backflow of information from the environment to the system providing a unique signature \cite{breuerN,alonso}.

\color{black}In recent times, much effort has been devoted to the study of quantum non-Markovian dynamics which provides advantages in several quantum information processing tasks such as perfect teleportation with mixed states \citep{task1}, efficient work extraction from Otto cycle \citep{task4}, efficient quantum control \citep{task5}, entangled state preparation \citep{task2,task3}, quantum metrology \citep{altherr2021quantum}, quantum evolution speedup \citep{deffner2017geometric}, and so on. Experimental realization of non-Markovianity has  been achieved in trapped-ion, nuclear magnetic resonance and photonic systems indicating a potential resource for executing quantum information processing tasks in real systems \cite{experiment1,experiment2,experiment3,experiment4,experiment6,experiment7,experiment8,experiment9,experiment11}.

Despite several interesting applications of non-Markovianity, a fundamental and important question is to assess whether the underlying quantum dynamics is non-Markovian at all, so that one can utilize it as a resource in legitimate quantum information processing tasks. Therefore, identifying whether a dynamics provides non-Markovian traits is a substantial task for  advancement of quantum technologies. Several methods have been proposed till date from different perspectives and utilizing different properties of non-Markovian dynamics \cite{RHP,blp1,bellomo,arend,Bhattacharya17,samya2,Bhattacharya20,Maity20,BBhattacharya21,Maity22,wolf2008assessing,jeknic2023invertibility}. In this work, we provide an adequate technique to efficiently detect  non-Markovian dynamics entailed with  environmental memory. Our 
approach is based on the determination of partial moments
of the Choi state, and does not require  full process tomography, thereby
making it easier to realize in a real experiment.

Our proposal utilizing the moment criterion  requires evaluation of simple functionals which can be efficiently estimated using an  experimental technique called shadow tomography \cite{aaronson2018shadow,aaronson2019gentle,Huang2020}.
It is based on a recently proposed methodology for simultaneous evaluation of several quantities for Noisy Intermediate Scale Quantum (NISQ) devices and is  more efficient than usual tomography. Moreover, it may be noted here that our criterion is state independent unlike the witness based detection scheme for which prior information about the quantum state is necessary. We further 
provide two explicit  examples in  support of our detection scheme for
non-Markovian evolution. 

In addition to the task of detecting a non-Markovian dynamics, another important task is to provide a quantitative measure of non-Markovianity. However, non-Markovianity  can be manifested in several ways indicating that there exists no common or general way of comparing  non-Markovian dynamics for different physical models. Two   measures of non-Markovianity proposed earlier, are based on the concept of divisibility of the dynamical map (RHP measure) \cite{RHP,rivas1,hakoshima2021relationship} and distinguishability of quantum states (BLP measure) \cite{BLP,blp1,breuerN}. In this work, we   define a measure based on partial moments of the Choi-matrix to quantify non-Markovianity.

The paper is organized as follows. In section \ref{s2}, we provide a brief
overview of the essential mathematical preliminaries concerning the dynamics of open quantum systems, as well as the moment criteria proposed in earlier works for entanglement detection. In  section \ref{s3} we present our  framework for detection of non-Markovianity  along with some explicit examples. A  measure of non-Markovianity is proposed in section \ref{s4}
where we also compare our proposed measure with the RHP measure for pure dephasing channel with Ohmic spectral density. Finally, in section \ref{s5}, we summarize our main findings.

\section{Preliminaries}\label{s2}
\subsection{Dynamics of Open Quantum System}
Isolated systems undergo unitary evolution. However, a general quantum evolution (or a quantum channel) can be represented by a completely-positive trace-reserving (CPTP) map $(\Lambda (t,t_0))$ which maps an element ($\rho (t_0)$) of the set of density operators \color{black}($B(\mathcal{H})$) to another element of the set i.e,  $\Lambda (t,t_0) : \rho (t_0) \mapsto \rho (t)$. The set of all such CPTP maps can be represented as $D$. We assume that the inverse $\Lambda^{-1} (t,t_0)$ exists for all time from $t_0$ to $t$. One can thus write the dynamical map for any $t \ge s \ge t_0$, into a composition 
\begin{equation}
\Lambda (t,t_0)= \Lambda (t,s) \circ \Lambda (s,t_0).  \label{divisibility}
\end{equation}
Even though $\Lambda (t,t_0)$ is always completely positive since it must correspond to a physically legitimate dynamics \color{black} (and hence $\Lambda^{-1} (t,t_0)$ is well defined) and $\Lambda (s,t_0)$ is completely positive,  the map $\Lambda (t,s)$ however, need not be completely positive. A dynamics acting on the system of interest is said to be \textit{divisible} iff it can be written as Eq. \eqref{divisibility} for any time $t \geq s \geq t_0, $ where $t_0$ is the initial time of dynamics, and $\circ$ represents the composition between two maps. The dynamics $\Lambda (t,t_0)$ is said to be \textit{positive divisible} (P divisible) if $\Lambda (t,s)$ is a positive map for every $t \geq s \geq t_0 $ satisfying the composition law. The dynamics $\Lambda (t,t_0)$ is said to be \textit{completely positive divisible} (CP divisible) if $\Lambda (t,s)$ is a CPTP map for every $t \geq s \geq t_0$ and satisfies the composition law.

The above mathematical characterization of a dynamical map $\Lambda (t,t_0)$ in terms of `divisibility', describing a memoryless evolution as a composition of physical maps leads to the definition of quantum Markovianity.  According to 
the RHP criterion, a dynamics is said to be non-Markovian if it is not CP-divisible \cite{RHP}. Another  way of characterizing non-Markovian dynamics is provided by Breuer {\it et. al.} \cite{BLP,blp1} where distinguishability of quantum states after the action of dynamical map is considered. Due to the interaction of a quantum system with the noisy environment, two quantum states lose their state distinguishability gradually with time. However, if at any instant of time, the distinguishability increases, then there is backflow of information from the environment to the system leading to the signature of non-Markovianity. The former way of representing a dynamics to be non-Markovian is known as RHP-type non-Markovianity \cite{RHP,rivas1}, whereas the latter one is known as BLP-type non-Markovianity \cite{BLP,breuerN}. A dynamics which is Markovian in the RHP sense is also Markovian in the BLP sense, but the converse is not true in general. Therefore, CP divisibility breaking is a necessary but not sufficient condition for information backflow from the environment to the system. In this paper we adopt CP divisibility as the sole property of quantum Markovianity, and any deviation from CP-divisibility (indivisible) will be considered as the benchmark of non-Markovianity.

Now, for each of the dynamical maps $\Lambda (t,t_0) \in D$, one can find a one-to-one correspondence to a state, called the Choi-state $\mathcal{C}_{\Lambda} (t,t_0) \in \mathcal{F}$ (where $\mathcal{F}$ is the set of all Choi-states) via channel-state duality where the Choi state \cite{choi} is defined as
\begin{equation}
  \mathcal{C}_{\Lambda} (t,t_0) = (\mathbb{I} \otimes \Lambda (t,t_0)) \ket{\phi}\bra{\phi} \label{choi}
\end{equation}
with $\ket{\phi}\bra{\phi}$ being a maximally entangled bipartite state of dimension $d \cross d$. According to the Choi–Jamiolkowski isomorphism \cite{choi,jamil}, for checking complete-positivity of $\Lambda (t,t_0)$, it is sufficient to check the positive semidefiniteness of the corresponding Choi-state  $(\mathcal{C}_{\Lambda} (t,t_0))$.

\subsection{Partial moment criterion}
In the bipartite scenario, one of the most well-known detection schemes of entanglement is based on the PPT criterion which examines whether the partial transposed state $\rho^{T_A}_{AB}$ (where partial transposition is taken w.r.t subsystem A) is positive semi-definite (all eigenvalues are non-negative) or not. Violation of this criterion implies that the given state $\rho_{AB}$ is entangled. This criterion has been shown to be a necessary and sufficient condition for $2\otimes 2$, $2\otimes 3$ and $3\otimes 2$ systems and has many applications in theoretical works \cite{castelnovo2013negativity,eisler2014entanglement,wen2016topological, ruggiero2016entanglement,blondeau2016universal,ruggiero2016negativity}. 
However, the transposition map not being a physical one, is impossible to implement exactly in an experimental scenario. A useful measure using this PPT criterion is the \textit{negativity} measure \cite{vidal2002computable}, defined as:
 \begin{equation}
 \mathbb{N}= ||\rho^{T_A}_{AB}|| = \sum_{i}{} |\lambda_{i}|  \nonumber
 \end{equation}
 where $\lambda_{i}$'s are the negative eigenvalues of $\rho^{T_A}_{AB}$.
 But this again requires an access to the full spectrum of $\rho^{T_A}_{AB}$, which is not obtainable through an experimental setup. To overcome this issue, the idea of moments of the partially transposed density matrix (PT-moments) was introduced to study the correlations in many-body systems in relativistic quantum field theory by Calabrese {\it et al.} in 2012 \cite{calabrese2012entanglement}.
 
  For a bipartite state $\rho_{AB}$, these PT-moments are given by,
\begin{equation}
P_n = \text{Tr}[{(\rho_{AB}}^{T_A})^n] \label{PTmoments}
\end{equation}
for n=1,2,3,.... One may note that $P_1 = \text{Tr}[\rho_{AB}^{T_A}] = 1$ while $P_2 = \text{Tr}[(\rho_{AB}^{T_A})^2]$  is related to the purity of the state. Therefore, $P_3$ is the first non-trivial moment which is necessary to capture additional
 information related to the partial transposition. Using only these first three PT moments, a simple but powerful entanglement detection criterion was proposed in ref. \cite{elben2020mixed}. This suggests that if a state $\rho_{AB}$ is PPT, then ${P_2}^2 \leq P_3 P_1 $. Therefore, from the contrapositivity of this statement, it follows that if a state $\rho_{AB}$ violates this condition, then it must be entangled which is the $p_3$-PPT criterion. Just like the PPT condition, this $p_3$-PPT condition is also applicable to mixed states and is a state independent criterion unlike entanglement witness \cite{Horodecki09,GUHNE09}. While entanglement witness provides a  stronger criterion for entanglement detection,  some prior knowledge about the state is required for the implementation of entanglement witness. 
 
 Even though this $p_3$-PPT criterion is weaker than the general PPT criterion, 
the former involves simple functionals which are easy to realise in a real experiment by a method called shadow tomography \cite{aaronson2018shadow,aaronson2019gentle,Huang2020}. For Werner states, the $p_3$-PPT criterion and the full PPT criterion are equivalent, and hence, the $p_3$-PPT criterion is a necessary and sufficient criteria for bipartite entanglement of Werner states. The PT-moments can be obtained experimentally with the help of shadow tomography without actually performing full state tomography, thus making it more efficient in terms of resources consumed. For a detailed discussion on shadow tomography and its advantage over general tomography, interested readers are referred to Refs. \cite{aaronson2018shadow,aaronson2019gentle,Huang2020,elben2020mixed}.
The technique of PT moments offers unparalleled advantage in NISQ and in many-body systems where a single qubit is used as a control and many distinct PT moments can be estimated from the same data unlike using random global unitaries for randomized measurements \cite{Huang2020,elben2020mixed}. Furthermore, the $p_3$-moment, in addition to detecting mixed state entanglement \cite{elben2020mixed,Neven2021, yu2021optimal} is also used to study entanglement dynamics in many-body quantum systems \cite{Brydges2019,Elben20}.

It might be noted here that the $p_3$-PPT condition provides a necessary condition for separability. However, each higher order moment ($n \geq 4$) gives rise to an independent and different entanglement detection criterion, and evaluating all the higher order moments provides a necessary and sufficient criterion for NPT entanglement \cite{Neven2021}. But this is very challenging from an experimental point of view and hence, moments up to third order are  used in order to provide an entanglement detection criteria and this simplifies the task.
Motivated by the above considerations, in the next section we explore whether a moment based detection scheme can be developed for non-Markovian dynamics, since in realistic scenarios, many different categories of dynamics consist of non-Markovian memory. We define $\Lambda$-moments $(r_n)$, and based on it we develop a formalism for detection of non-Markovian dynamics characterized via indivisibility. 

\section{Detection of non-Markovianity} \label{s3}
\textbf{Definition 1:} Let $\Lambda (t,s)$ be a trace preserving, linear map that satisfies the composition law \eqref{divisibility}. We define the $n$th order $\Lambda$-moments $(r_n) $ as:
\begin{equation}
r_n = \text{Tr}[ (\mathcal{C}_{\Lambda} (t,s))^n] \label{lambdamoments}
\end{equation}
with $n$ being an integer and $\mathcal{C}_{\Lambda} (t,s)$ is the Choi state (defined earlier) corresponding to the dynamics acting on the system between the time intervals $s$ and $t$, such that $s \le t$.
With the above definition, we are now ready to propose our criterion for detecting non-Markovian dynamics.

\textbf{Theorem 1:} If a dynamics is Markovian, then 
\begin{equation}
{r_2}^2 \leq r_3 \label{6}
\end{equation}
where $r_2$ and $r_3$ are defined in \eqref{lambdamoments}.

\proof  If $t_0$ be the initial time of a dynamics, then the time evolution of an open
quantum system is governed by a family of completely positive trace preserving maps $\{{\Lambda(t,t_0)\}}_{t \ge t_0}$ satisfying the  composition law \eqref{divisibility}. 
%The map $\Lambda (t,s)$ decides whether the nature of the dynamics is Markovian or non-Markovian.
Considering the concept of Choi-Jamiolkowski isomorphism \cite{choi,jamil}, we shall henceforth use the Choi operator ($\mathcal{C}_{\Lambda}$) corresponding to the map $\Lambda (t,s)$. 

Let us now consider the Schatten-$p$ norms for $p \ge 1$, which are defined as
 \begin{equation}
    ||X||_{p} = (\sum_{i=1}^{n}{|\chi_i|^p})^{\frac{1}{p}}=(\text{Tr}[|X|^p])^{\frac{1}{p}}   \label{schattenp}
\end{equation}
 where $X$ is a $n \times n$ hermitian matrix having eigenvalue decomposition $X= \sum_{i=1}^{n} \chi_{i} \ket{x_i}\bra{x_i}$. 
 Replacing $X$ by the Choi matrix $\mathcal{C}_{\Lambda}$ %corresponding to the map $\Lambda(t,s)$ 
 in Eq.\eqref{schattenp}, the Schatten-$p$ norms for Choi matrix are analogously defined.  Further, the $l_p$ norm of the vector of eigenvalues of $\mathcal{C}_{\Lambda}$ corresponding to each Schatten-$p$ norm is defined by:
  \begin{equation}
     || \lambda||_{l_p} := (\sum_{i=1}^{n}{|\lambda_i|^p})^{\frac{1}{p}} \label{lpnorm}
 \end{equation}
 where ${\{\lambda_i \}_{{i=1}}^{n}}$ is the spectrum of  $\mathcal{C}_{\Lambda}$.
 The inner product corresponding to an $n$ vector is defined as
 \begin{equation}
     \langle u, v \rangle :=\sum_{i=1}^{n} u_{i} v_{i} \label{innerproduct}
 \end{equation}
 for $u,v \in \mathbb{R}^{n}$.
 Now, from Hoelder's inequality for vector norms, we know that for $p, q \ge 1$ and $\frac{1}{p}+\frac{1}{q} =1,$
 the following relation holds:
 \begin{equation}
     |\langle u, v\rangle | \le \sum_{i=1}^{n} |u_{i} v_{i}| \le ||u||_{l_p} ||v||_{l_q} .\label{hoelderinequality}
 \end{equation}
 Putting $p=3$ and $q=\frac{3}{2}$ in \eqref{hoelderinequality}, we get
 \begin{equation}
     \text{Tr}[(\mathcal{C}_{\Lambda})^2]= \langle\lambda, \lambda\rangle \le ||\lambda||_{l_3} ||\lambda||_{l_{\frac{3}{2}}}=||\mathcal{C}_{\Lambda}||_{3} ||\lambda||_{l_{\frac{3}{2}}} \label{applyinghoelder}
 \end{equation}
 We next apply the Cauchy-Schwarz inequality which is obtained by putting $p=\frac{1}{2}$ and $q=\frac{1}{2}$ in Hoelder's inequality. Therefore, 
 \begin{align}
&{||\mathcal{C}_{\Lambda}||_{2}}^2=\text{Tr}[(\mathcal{C}_{\Lambda})^2] \nonumber \\ 
& ~~~~~ \leq ||\mathcal{C}_{\Lambda}||_{3} ||\lambda||_{l_{\frac{3}{2}}} \nonumber \\ 
&   ~~~~~ =  ||\mathcal{C}_{\Lambda}||_{3} (\sum_{i=1}^{n}{|\lambda_i|^{\frac{3}{2}}})^{\frac{2}{3}} \nonumber \\ 
&     ~~~~~ =  ||\mathcal{C}_{\Lambda}||_{3} (\sum_{i=1}^{n}{{|\lambda_i|}{|\lambda_i|}^{\frac{1}{2}}})^{\frac{2}{3}} \nonumber \\ 
&     ~~~~~  \leq  ||\mathcal{C}_{\Lambda}||_{3}  ((\sum_{i=1}^{n}{|\lambda_i|^2})^{\frac{1}{2}} (\sum_{i=1}^{n}{|\lambda_i|})^{\frac{1}{2}})^{\frac{2}{3}}\nonumber \\ 
&     ~~~~~ =||\mathcal{C}_{\Lambda}||_{3}  {||\mathcal{C}_{\Lambda}||_{2}}^{\frac{2}{3}} {||\mathcal{C}_{\Lambda}||_{1}}^{\frac{1}{3}}
 \label{applyingcauchy}
\end{align}
Taking $3$rd power of \eqref{applyingcauchy}, we get
\begin{equation}
   {||\mathcal{C}_{\Lambda}||_{2}}^4 \le  {||\mathcal{C}_{\Lambda}||_{3}}^3 ||\mathcal{C}_{\Lambda}||_{1} \label{applying3rdpower}
\end{equation}
Since $\Lambda$ is a trace preserving map, $||\mathcal{C}_{\Lambda}||_{1} =1$, and hence \eqref{applying3rdpower} reduces to,
\begin{equation}
   {||\mathcal{C}_{\Lambda}||_{2}}^4 \le  {||\mathcal{C}_{\Lambda}||_{3}}^3  \label{normproof}
\end{equation}
i.e.,
\begin{equation}
   (\text{Tr}[(\mathcal{C}_{\Lambda})^2])^2 \le \text{Tr}[(\mathcal{C}_{\Lambda})^3] \label{theorem1proof}
\end{equation}
which completes the proof. \qed

The above theorem indicates that condition \eqref{6} is necessary for a dynamics to be Markovian. Violation of above theorem is therefore sufficient to conclude that the underlying dynamics is actually CP-indivisible and hence non-Markovian. Below, we present some explicit examples of non-Markovian dynamics which can be detected by the condition mentioned above.

\subsection{Examples:} We would now like to present two examples in support of Theorem 1.  It may be noted that here we will consider the set of operations which have Lindblad type generators. For system density matrix $\rho$, the Lindblad master equation can be written as 
\begin{equation}
    \frac{d\rho}{dt} = -\frac{\iota}{\hbar} [H, \rho] + \sum_{i} \gamma_i (L_i \rho {L_i}^{\dagger} - \frac{1}{2}({L_i}^{\dagger}  L_i \rho + \rho {L_i}^{\dagger}  L_i ))
\end{equation}
where the unitary aspects of the dynamics is described by the Hamiltonian $H$, $\gamma_i$ are the Lindblad coefficients, and $L_i$ are the Lindblad operators which describe the dissipative part of the dynamics \cite{lindblad}. \\

\textit{Example 1}: 
We consider a qubit system  that interacts with an amplitude damping environment which is modeled by another qubit system. The non-Markovian character of this model studied earlier \cite{mukherjee2015efficiency}, 
is motivated by the experimental realization of such non-Markovian dynamics
through the violation of temporal Bell-like inequalities in a controllable Nuclear Magnetic Resonance system \cite{souza2013experimental} . 

\begin{figure}[ht]
\includegraphics[width=.45\textwidth]{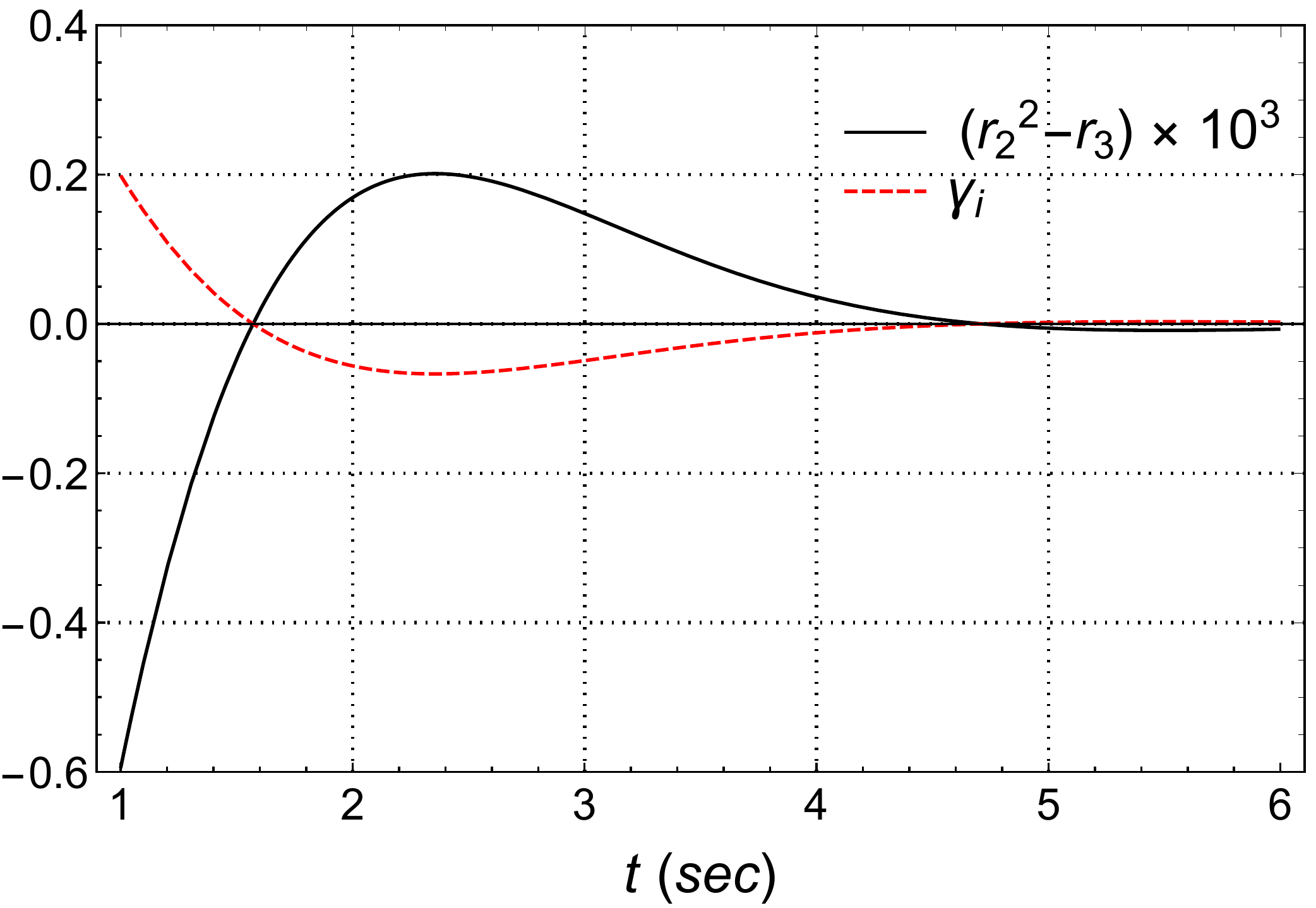}
\caption{Non-Markovian behavior of the dynamics given by Eq.\eqref{example1}, ${r}_{2}^2 - {r}_{3} > 0$, is exhibited  for $\gamma <0$.}\label{fig1}
\centering
\end{figure}

The master equation is  given by
\begin{equation}
\frac{d\rho}{dt} = \mathcal{L} (\rho)=  \gamma_1(\sigma_{x} \rho  \sigma_{x}-\rho)+ \gamma_2(\sigma_{y} \rho  \sigma_{y}-\rho)+\gamma_3(\sigma_{z} \rho  \sigma_{z}-\rho) \label{example1}
\end{equation}
 The Lindblad coefficients are taken as $\gamma_1=\gamma_2=\gamma_3= e^{-t'} \cos{t'}$ (for all $i$) and $t'=kt$, with $k$ being a constant having the dimension of $[T^{-1}]$.
 The corresponding dynamical map is given by $\Lambda (\rho) = e^{\mathcal{L}}(\rho)$. For small time approximation (i.e, $|\epsilon \gamma_i| << 1$), the Choi state corresponding to $\Lambda (\rho)$ is given by  
\begin{equation}
\mathcal{C}_{\Lambda} = (\mathbb{I} \otimes (\mathbb{I}+\epsilon \mathcal{L})) \ket{\phi^+} \bra{\phi^+} \label{choilindblad}
\end{equation}
with $\ket{\phi^+}=\frac{\ket{00}+\ket{11}}{\sqrt{2}}$. Here, we consider $\epsilon=0.001$ and $k=1$.
It is known that the above dynamics shows its non-Markovian nature when $\gamma_i <0$. Therefore, for $\gamma_i <0$, we should have  ${r}_{2}^2 - {r}_{3} > 0$, which is evident from Fig. \ref{fig1}.

\vspace{5mm}

\textit{Example 2}: As a second example, we consider a pure dephasing non-Markovian dynamics. A qubit system interacts with a thermal reservoir which  is modeled by an infinite set of harmonic oscillators in the vacuum state. The Hamiltonian corresponding to the system-reservoir interaction is 
\begin{equation}
H=\omega_{\sigma}\sigma_{z}+\sum_{k}\omega_{k}a_{k}^{\dagger}a_{k}+\sum_{k}\alpha_{k}\sigma_{z}[a_{k}exp(i\theta_{k})\\+a_{k}^{\dagger}exp(-i\theta_{k})] \nonumber
\end{equation}
where $\omega_{\sigma},\omega_{k}$ are the energy gap of the system and the frequency of $k$-th mode of reservoir, respectively, $a_{k}^{\dagger}(a_{k})$ are creation (annihilation) operators of the harmonic oscillator, $\alpha_{k}$ is the coupling constant for the $k$-th mode, and $\theta_{k}$ is the corresponding phase. The master equation is  given by 
\begin{equation}
\frac{d\rho}{dt}=\mathcal{L}(\rho)=\gamma (\sigma_{z}\rho(t)\sigma_{z}-\rho(t)) \label{example2}
\end{equation} 
The time dependent decay rate $\gamma$ corresponding to a Lorentzian spectral density is \citep{salimi2016role,mukherjee2015efficiency}
\begin{equation}
    \gamma = \frac{2\lambda \gamma_{0} \sinh(t'g/2)}{g\cosh(t'g/2)+\lambda\sinh(t'g/2)}\label{g}
\end{equation}
 with $g= \sqrt{{\lambda}^2 - 2\gamma_0 \lambda}$, $t'=kt$. Here $k$ is a constant having the dimension of $[T^{-1}]$, $\lambda$ is the spectral width and $\gamma_{0}$ is the coupling strength. 
 
\begin{figure}[ht]
\includegraphics[width=.45\textwidth]{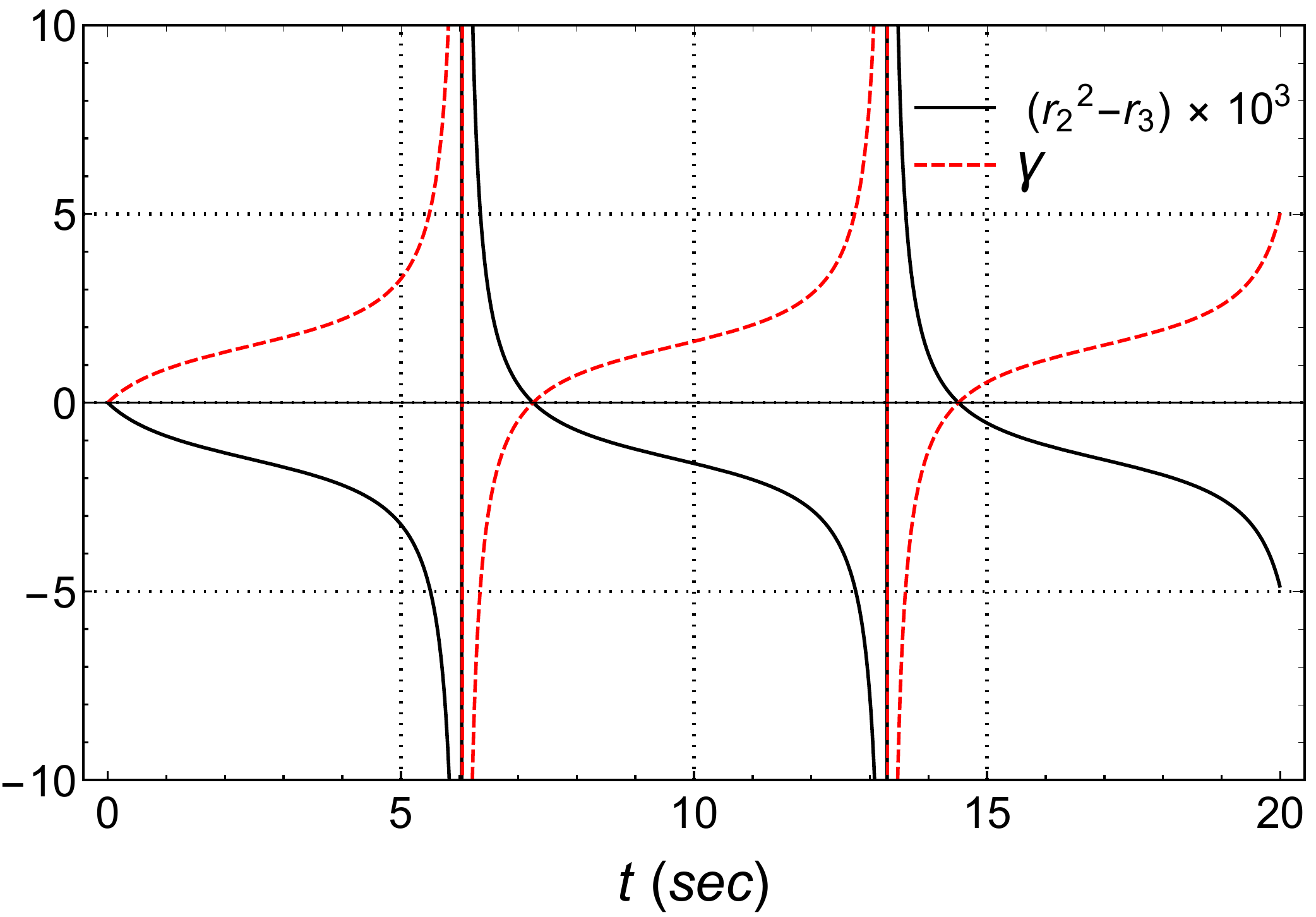}
\caption{Non-Markovian behavior of the dynamics given by Eq.\eqref{example2}, ${r}_{2}^2 - {r}_{3} > 0$, is exhibited  for $\gamma <0$.  }\label{fig2}
\centering
\end{figure}

In the small time approximation (i.e, $|\epsilon \gamma | << 1$), the Choi state corresponding to the dynamical map, $\Lambda (\rho) = e^{\mathcal{L}}(\rho)$ is given by $\mathcal{C}_{\Lambda} = (\mathbb{I} \otimes (\mathbb{I}+\epsilon \mathcal{L})) \ket{\phi^+} \bra{\phi^+}$ where $\ket{\phi^+}$ is the maximally entangled state defined earlier.
It is known that the above dynamics shows its non-Markovian nature when $\gamma <0$ which is possible only when $\gamma_0 > \lambda /2$ \cite{mukherjee2015efficiency}. So, for $\gamma <0$, we should have ${r}_{2}^2 - {r}_{3} > 0$ which is again evident from Fig. \ref{fig2}. We consider here $\lambda = 1.5$, $\gamma_0 =1$ and $k=1$ for the figure.

\section{Measure of non-Markovianity}  \label{s4}
In this section, we would like to define a quantitative measure of non-Markovianity. Using Schatten-$p$ norms for $p=2$ and $p=3$ we define a measure of non-Markovianity. Let us first denote:
\begin{equation}
f(t)= \lim_{ \epsilon \to 0} \frac{M_{\epsilon}}{\epsilon } \label{measure}
\end{equation}
where,
\begin{equation}
\begin{split}
      M_{\epsilon}&= {M_{T}}({t+\epsilon},t) \\ &=
    \text{max} \{0,({||\mathcal{C}_{\Lambda} (t+\epsilon,t)||_{2}}^2)^{2} -{||\mathcal{C}_{\Lambda} (t+\epsilon, t)||_{3}}^3 \} \label{value of measure}
    \end{split}
\end{equation}
   We define,
 \begin{equation}
     \mathcal{M}=  \int_{0}^{\infty} f(t) \,dt \label{our measure}
 \end{equation} as a measure of non-Markovianity.
 
Below we will show that $\mathcal{M}$ can be used as a measure of non-Markovianity for unital dynamics.
To show that $\mathcal{M}$ as a measure of non-Markovianity for unital dynamics, we need to show that $\mathcal{M} = 0$ for all Markovian dynamics and $\mathcal{M}$ is monotone under divisible unital dynamics.  It has been shown earlier that for all unital dynamical maps having corresponding Lindblad generators, the Lindblad operators are normal \cite{Bhattacharya2020}. Therefore, to prove the monotonicity of our measure we will consider that the Lindblad operators are normal.

\textbf{Lemma 1:} If the $\alpha$-Renyi entropy defined by  
$$S_{\alpha} (t) := \frac{1}{1-\alpha} \log _{2} \text{Tr} [X^{\alpha} (t)], ~~ (\alpha > 0)$$ evolves under Lindblad type divisible operation having normal Lindblad operator, then 
\begin{equation}
    \frac{d}{dt} S_{\alpha} (t) \geq 0. \label{monotoneofrenyi}
\end{equation}

\proof In Refs. \cite{Abe2016,Beigi2013,Hongting17,Bhattacharya2020, benatti1988entropy, das2018fundamental}, it has been shown that 
\begin{equation}
    \frac{d}{dt} S_{\alpha} (t) = 2 \sum_{i} \gamma_{i} (t) \chi_{i} (t) \label{renyievolution}
\end{equation}
where $$\chi_{i} (t) = \frac{\alpha}{1-\alpha} \frac{1}{\text{Tr}[X^{\alpha} (t)]}\text{Tr}[X^{\alpha -1} (t)L_{i}X(t)L_{i}^{\dagger} - X^{\alpha } (t)L_i^{\dagger} L_i ]$$ with $L_i$ being the Lindblad operator and $\gamma_i$ are the Lindblad coefficients. Further \cite{Abe2016},
\begin{equation}
    \chi_{i} (t) > \langle [ L_i^{\dagger}, L_i ]\rangle_{\alpha}  \label{normallindblad}
\end{equation}
where, $\langle A \rangle_{\alpha} = \text{Tr}[A X^{\alpha } (t)]/\text{Tr}[ X^{\alpha } (t)]$. For normal Lindblad operator $[ L_i^{\dagger}, L_i ] =0$. Therefore, from \eqref{normallindblad}, $\chi_{i} (t) > 0$.  Also, if the dynamics is divisible, i.e., all $\gamma_i >0$, then from Eq. \eqref{renyievolution} it immediately follows that $ \frac{d}{dt} S_{\alpha} (t) \geq 0$. \qed 

\vspace{1.4mm}
\textbf{Lemma 2:} If the Schatten-$p$ norms  of a positive, Hermitian operator $X$ evolve under divisible operation having Lindblad type generators, and the Lindblad operators are normal, then
\begin{equation}
    ||X{(t+\delta t)}||_p \leq ||X(t)||_p .\label{lemma2}
\end{equation}
where $||X{(t+\delta t)}||_p,||X(t)||_p $ are the Schatten-$p$ norms of $X$ at times $t+\delta t$ and $t$, respectively.

\proof 
 From Lemma 1 it follows   that for Lindblad type divisible operation having normal Lindblad operator, \begin{align}
    &~~~~~~~~~~S_{\alpha} (t+ \delta t) - S_{\alpha} (t) \geq 0 \nonumber \\
    &\implies \frac{1}{1-\alpha} \log_{2} \text{Tr} [X^{\alpha} (t +\delta t)] \geq \frac{1}{1-\alpha} \log_{2} \text{Tr} [X^{\alpha} (t)]. \nonumber \\
    &\text{Now, if $\alpha > 1$, then $\frac{1}{1-\alpha} <0$ and hence} \nonumber \\
    &~~~~~~~~ - \log_{2} \text{Tr} [X^{\alpha} (t +\delta t)] \geq - \log_{2} \text{Tr} [X^{\alpha} (t)]  \nonumber \\
   & \implies \alpha \log_{2} \text{Tr} [X^{\alpha} (t +\delta t)] \leq \alpha \log_{2} \text{Tr} [X^{\alpha} (t)] \nonumber \\
   & \implies \text{Tr} [X^{\alpha} (t +\delta t)]^{\alpha } \leq \text{Tr} [X^{\alpha} (t)]^{\alpha } .\label{traceformoflemma2}
\end{align}

Since $\alpha >0 $, by taking the ${\alpha}$-th root of both sides, we get 
  \begin{align}
      \text{Tr} [X^{\alpha} (t +\delta t)] \leq  \text{Tr} [X^{\alpha} (t)].
      \end{align}
From definition of  Schatten-$p$ norms, it follows
      \begin{align}
{||X{(t+\delta t)}||_{\alpha}}^{\alpha} \leq {||X(t)||_{\alpha}}^{\alpha} .
  \end{align}

  Again,  taking the ${\alpha}$-th root of both sides, we get,
  \begin{align}
       {||X{(t+\delta t)}||_{\alpha}} \leq {||X(t)||_{\alpha}} 
  \end{align}
  Now, replacing $\alpha$ by $p$, it follows that
   \begin{align}
       {||X{(t+\delta t)}||_{p}} \leq {||X(t)||_{p}} 
  \end{align} \label{traceformoflemma2}
  \qed
\vspace{1.4mm}

The above two lemmas  imply the following proposition.

\vspace{0.25cm}

\textbf{Proposition 1:} For unital dynamics having Lindblad type generators, $\mathcal{M}$ is monotone under divisible operation.

\proof In order to show $\mathcal{M}$ to be {\it monotone} under unital divisible dynamics, we have to prove that $\mathcal{M}$ satisfies 
 \begin{equation}
 \begin{split}
   ( \text{Tr}[({\mathcal{C}_{\Lambda}{(t+\delta t)})^2}])^2  -  \text{Tr}[({\mathcal{C}_{\Lambda}{(t+\delta t)}})^3] & \\  \leq ( \text{Tr}[({\mathcal{C}_{\Lambda}{(t)})^2}])^2 -  \text{Tr}[({\mathcal{C}_{\Lambda}{(t)}})^3]  \label{proposition1}
 \end{split}
\end{equation} 
 It is known that for unital dynamics having Lindblad type generators, the Lindblad operators are normal \cite{Bhattacharya2020}. 
  Therefore, replacing $X$ by $\mathcal{C}_{\Lambda}$, lemma 2 implies that for $p=2$ (it follows from \eqref{lemma2}),
%Now, replacing $X$ by $\mathcal{C}_{\Lambda}$, lemma 2 implies that for $p=2$ (it follows from \eqref{lemma2}),
\begin{equation}
     ||\mathcal{C}_{\Lambda}{(t+\delta t)}||{_2}^{4} \leq ||\mathcal{C}_{\Lambda}{(t)}||{_2}^{4} .\label{lemma2choi}
\end{equation}
Using  \eqref{lemma2choi}, we get
 \begin{equation}
 \begin{split}
    ||\mathcal{C}_{\Lambda}{(t+\delta t)}||{_2}^{4}-  ||\mathcal{C}_{\Lambda}{(t+\delta t)}||{_3}^{3}  & \\=  ( \text{Tr}[({\mathcal{C}_{\Lambda}{(t+\delta t)})^2}])^2 -  \text{Tr}[({\mathcal{C}_{\Lambda}{(t+\delta t)}})^3] & \\ \leq  ( \text{Tr}[({\mathcal{C}_{\Lambda}{(t)})^2}])^2 -  \text{Tr}[({\mathcal{C}_{\Lambda}{(t+\delta t)}})^3]  \label{applying2norm}
 \end{split}
 \end{equation}
For any quantum state $\rho$, evolving under the Lindblad type generators $\mathcal{L}$, 
\begin{equation}
\frac{d\rho}{dt} = e^{\mathcal {L}} \rho    \label{rhoevolution}
\end{equation}
it follows from \eqref{applying2norm} that,
\begin{align}
    & ( \text{Tr}[({\mathcal{C}_{\Lambda}{(t+\delta t)})^2}])^2 -  \text{Tr}[({\mathcal{C}_{\Lambda}{(t + \delta t)}})^3] \nonumber \\ 
 &   ~~~~~ \leq  ( \text{Tr}[({\mathcal{C}_{\Lambda}{(t)})^2}])^2 -  \text{Tr}[{(I+\epsilon \mathcal{L}+{\epsilon}^2{\mathcal{L}}^2 + ...)({\mathcal{C}_{\Lambda}(t)})^3}] \nonumber \\ 
 &     ~~~~~ \leq ( \text{Tr}[({\mathcal{C}_{\Lambda}{(t)})^2}])^2 -  \text{Tr}[({\mathcal{C}_{\Lambda}{(t)}})^3] \label{proposition1proofcomplete}
\end{align}

 Using \eqref{proposition1proofcomplete}, it can be seen that, 
\begin{equation}
    \mathcal{M}(t+\delta t) \leq \mathcal{M}(t)
\end{equation}
which proves that $\mathcal{M}$ is monotone under divisible unital dynamics.
This completes the proof. \qed

\vspace{1.4mm}

\textbf{Theorem 2:} $\mathcal{M}$ is a measure of unital Lindblad type non-Markovian dynamics having normal Lindblad operators.

\proof By the definition of $\mathcal{M}$ and from the proof of Theorem 1, it is clear that $\mathcal{M} =0$ for all Markovian dynamics. Furthermore, from Proposition 1 it follows that $\mathcal{M}$ is monotone under divisible operation for Lindblad type dynamics having normal Lindblad operators. Therefore, $\mathcal{M}$ satisfies  the necessary properties of a measure, and can be taken to be a measure of non-Markovianity for unital dynamics. \qed 
\\

\subsection{Comparison between our Measure and the RHP Measure}%by considering Amplitude damping channels with Lorentzian spectral density}
For a detailed comparison between our proposed measure ($\mathcal{M}$) and the RHP measure ($\mathcal{I}$), we consider pure dephasing channel as presented in \textit{Example 2}. The time dependent dephasing rate is given by \cite{salimi2016role}
\begin{equation}
    \gamma(t) = \int \frac{J(\omega) \coth({\hbar\omega / 2 k_{B} T}) \sin(\omega t)}{\omega} d\omega
\end{equation}
where, $\omega$ and $J(\omega)$ represents the frequency of the reservoir modes and spectral function of the reservoir respectively. For Ohmic spectral density, the spectral function is given by 
\begin{equation}
    J(\omega) = \omega \exp{\frac{-\omega}{\omega_{c}}}
\end{equation}
where, $\omega_{c}$ is the cut-off frequency of the reservoir.

In the small time approximation (i.e, $|\epsilon \gamma | << 1$), the Choi state corresponding to the dynamical map, $\Lambda (\rho) = e^{\mathcal{L}}(\rho)$ is given by $\mathcal{C}_{\Lambda} = (\mathbb{I} \otimes (\mathbb{I}+\epsilon \mathcal{L})) \ket{\phi^+} \bra{\phi^+}$ where $\ket{\phi^+}$ is the maximally entangled state.\\

\textbf{\textit{RHP Measure:}} From the Choi–Jamiolkowski isomorphism, we know that the dynamical map $\Lambda_{\epsilon} = \Lambda (t + \epsilon, t)$ is CP iff $(\mathbb{I} \otimes \Lambda_{\epsilon})  \ket{\phi^+} \bra{\phi^+} \ge 0$ $\forall \epsilon$. From the trace preserving condition, $||(\mathbb{I} \otimes \Lambda_{\epsilon})  \ket{\phi^+} \bra{\phi^+}||_{1} = 1$ iff $\Lambda_{\epsilon}$ is completely positive and  $||(\mathbb{I} \otimes \Lambda_{\epsilon})  \ket{\phi^+} \bra{\phi^+}||_{1} > 1$, if $\Lambda_{\epsilon}$ is not completely positive. Considering these facts, one can define
 \begin{equation}
     g(t) = \lim_{ \epsilon \to 0} \frac{||(\mathbb{I} \otimes \Lambda _{\epsilon}) \ket{\phi}\bra{\phi}||_{1}-1}{\epsilon}
 \end{equation}
 where $\Lambda _{\epsilon}= (\mathbb{I}+\epsilon \mathcal{L})$.
 It is clear that $g(t) \ge 0$ and $g(t) =0$ iff $\Lambda _{\epsilon}$ is CP. The integral
 \begin{equation}
      \mathcal{I}= \int_{0}^{\infty} g(t) \,dt
 \end{equation}
 is the RHP measure \cite{rivas1,RHP}.\\

 In case of pure dephasing channel with Ohmic spectral density, $\mathcal{C}_{\Lambda} = (\mathbb{I} \otimes (\mathbb{I}+\epsilon \mathcal{L})) \ket{\phi^+} \bra{\phi^+}$ has eigen values
   $\lambda_1= 0$, $\lambda_2=0$, $\lambda_3= \gamma \epsilon$, $\lambda_4= 1- \gamma \epsilon$.
 \\
 Therefore,
 \begin{align}
      g(t)=
     \begin{cases}0 \hspace{1cm} \text{when} \hspace{0.2cm}\gamma(t) \ge 0\\
   -2 \gamma, \hspace{0.4cm}  \text{when} \hspace{0.2cm} \gamma(t)<0. \label{RHP measure}
    \end{cases}
 \end{align}
and 
\begin{equation}
    \mathcal{I} = \int_{\gamma < 0}  -2 \gamma  \,dt \label{final RHP measure}
\end{equation}

    \textbf{\textit{Our Proposed Measure:}} For the same example, the value of  $f(t)$ defined in Eq.(\ref{measure}) turns out to be 
     \begin{align}
 f(t) = 
     \begin{cases}0 , ~~~~~~~~~~~~~~~~\text{when} \hspace{0.2cm}\gamma(t) \ge 0\\
   - \gamma , \hspace{0.8cm} ~~~ \text{when} \hspace{0.2cm} \gamma(t)<0.\label{our measure}
    \end{cases}
    \end{align}
    Therefore, our proposed measure becomes
    \begin{equation}
         \mathcal{M} = \int_{\gamma < 0}  - \gamma  \,dt .\label{final our measure}
    \end{equation}

Comparing Eqs.\eqref{final RHP measure} and \eqref{final our measure}, it turns out that the two measures are related by $\mathcal{I} = 2 \mathcal{M}$.

It is worth noting that a straightforward calculation for other well-known channels, such as the depolarizing channel, reveals that the RHP measure is twice that of our proposed measure. However, establishing a general relationship between the two measures for arbitrary channels remains a subject of further investigation.

\color{black}\section{Conclusions}\label{s5} 

Non-Markovianity of open quantum system dynamics has already been established as a useful resource for several information processing tasks \cite{task1,task2,task3,task4,task5}. However, prior to incorporating it into any information processing task, it is of foremost importance to detect the signature of such a resource. In this work, we have developed a methodology to assess whether a dynamics attributes non-Markovian traits. Our proposal is based on the moment criterion of Choi matrices, which can be efficiently demonstrated in an experimental setup. We have presented two explicit examples in order to illustrate our detection scheme. Further, we have proposed a measure of non-Markovianity for unital dynamics which is again based on the partial moment criterion. Interestingly, our proposed measure turns out to be just half of the RHP measure for some of the well-known channels. However, the exploration of a general relationship between these two measures for arbitrary channels requires further analysis.

Our proposed non-Markovianity detection protocol requires computation of simple functionals without necessitating the evaluation of full spectrum of the evolution, and hence, is a lot more efficient than full process tomography \cite{elben2020mixed}. Moreover, the protocol being state independent, no prior information about the dynamics is required unlike witness based detection schemes \cite{Huang2020}. Furthermore, since our proposed measure relies on moments of the Choi state, it is amenable for implementation in experiments utilizing the tools of shadow tomography\cite{Huang2020}. The relevance  of our proposed detection criterion should be further evident in the case of non-Markovian dynamics involving multi-qubit systems wherein an exponentially lesser number of samples would be required. Extension of
the moments based criterion for non-Markovianity detection of multi-qubit systems is therefore motivated as a natural off-shoot of our present analysis.  

\section{Acknowledgements}
BM and SM acknowledge Bihalan Bhattacharya for fruitful discussions. BM acknowledges DST INSPIRE fellowship program for financial support. ASM acknowledges support from the Project No. DST/ICPS/QuEST/2018/98 of the Department of Science and Technology, Government of India.
\bibliography{nonmarkovianity}

\end{document}